# Tailoring topological order and π-conjugation to engineer quasi-metallic polymers

B. Cirera,[1,†] A. Sánchez-Grande,[1,†] B. de la Torre,[2,3,†] J. Santos,[1,6] S. Edalatmanesh,[3] E. Rodríguez-Sánchez,[1,6] K. Lauwaet,[1] B. Mallada-Faes,[3] R. Zbořil,[3] R. Miranda,[1,5] O. Gröning,[4,*] P. Jelínek,[2,3]* N. Martín,[1,6]* D. Écija[1,*]

**Affiliations:**

[1] IMDEA Nanociencia. 28049, Madrid, Spain.

[2] Institute of Physics. The Czech Academy of Sciences, 162 00 Prague, Czech Republic.

[3] Regional Centre of Advanced Technologies and Materials. Palacký University, 78371 Olomouc, Czech Republic.

[4] nanotech@surfaces laboratory. EMPA - Swiss Federal Laboratories for Materials Science and Technology, 8600 Dübendorf, Switzerland

[5] Departamento de Física de la Materia Condensada. Facultad de Ciencias. Universidad Autónoma de Madrid, 28049 Madrid, Spain.

[6] Departamento de Química Orgánica. Facultad de Ciencias Químicas. Universidad Complutense de Madrid, 28040 Madrid, Spain.

*Correspondence to: oliver.groening@empa.ch (O.G), jelinekp@fzu.cz (P.J.), nazmar@quim.ucm.es (N.M.), david.ecija@imdea.org (D.E.)

† Equally contributing authors.

**Abstract:**

Topological band theory provides a conceptual framework to predict or even engineer robust metallic states at the boundaries of topologically distinct phases. The bulk-boundary correspondence requires that a topological electronic phase transition between two insulators must proceed via closing of the electronic gap. Therefore, it can provide a conceptual solution to the instability of metallic phases in π-conjugated 1D polymers.

In this work we predict and demonstrate that a clever design and on-surface synthesis of polymers consisting of 1D linearly bridged polyacene moieties, can position the resulting polymer near the topological transition from a trivial to a non-trivial quantum phase featuring a very narrow bandgap with in-gap zero-energy edge-states at the topologically non-trivial phase. We also reveal the fundamental connection between topological classes and electronic transformation of 1D π-conjugated polymers.

**One Sentence Summary:** The on-surface chemical reactions of molecular precursors on a pristine Au(111) surfaces results in the formation of long, defect-free, π-conjugated polymers featuring a topological non-trivial SSH quantum phase, with a very narrow bandgap due to the proximity to the point of topological transition, in-gap zero-energy edge-states and electronic transformation of the π-conjugated polymers.

**Main Text:**

The tremendous advances in characterizing and understanding materials at the atomic limit have made possible nowadays to engineer nanomaterials with highly specific electronic properties, spanning distinct domains such as molecular electronics, optoelectronics, spintronics and topology.

In this respect, π-conjugated polymers have received particular interest from scientists and industry alike, since the discovery of high conductivity in doped polyacetylene showed that it was possible to create organic synthetic metals (*1*). Such finding opened the new field of what is known today as organic electronics (*2*). At the same time intense research at the boundary between chemistry and condensed-matter physics was triggered, which contributed to progress in understanding the fundamental chemistry and physics of π-conjugated macromolecules (*3*). Notably, the relation between topological band theory and conduction in doped polyacetylene (PA) was soon introduced by the Su-Shrieffer-Heeger (SSH) model (*4*). Steered by this discovery, great efforts have been devoted during last decades to rationally relate chemical structure and electronic properties in these materials (*5*). In this regard, gaining control over the bandgap of π-conjugated polymers has been a target of particular interest (*6, 7*). As it depends mainly upon the molecular structure of the repeating unit, synthetic chemists have been provided with the opportunity and the challenge of controlling the bandgap by design at the molecular level, only precluded by synthetic limitations, such as solubility (*8*), whereby a great attention is devoted to reduce or even quench it. Narrow bandgap π-conjugated polymers are of utmost interest because of their near-infrared activity, high conductivity, and ambipolar charge transport character (*9-11*). Furthermore, the synthesis of intrinsic conductive π-conjugated polymers has remained elusive. One strategy towards this challenge is to design π-conjugated polymers with a small energy difference between the aromatic and quinoidal resonant structures simultaneously reducing the band gap (*12-15*). However, despite the efforts devoted, nowadays the only feasible way to enhance the conductivity of polymers (as for PA) is via doping (either chemically or electrochemically), although the process is known to be detrimental to their stability (*3*).

The rapid development of on-surface chemical synthesis has emerged as a new paradigm for engineering nanomaterials with atomic precision (*16-18*), while it also circumvents the solubility limitations associated to synthetic wet-chemistry. The might of this novel approach was recently exemplified by graphene nanoribbons (*19*), whereby robust topological SSH quantum phases were rationally engineered (*20, 21*). Although on-surface synthesis was up to now primarily concerned with graphene derived nanostructures, for future applications (*22*) it will be desirable to design long, transferable and defect-free nanomaterials, being π-conjugated polymers ideal candidates.

Following this paradigm, we have just reported the pathway to engineer surface-confined ethynylene-bridged anthracene polymers featuring high π-conjugation (*23*). Inspired by this recent discovery, we introduce here the theoretical basis and the experimental demonstration of the rational engineering of polymers that are topologically non trivial, featuring very narrow band gap and in-gap zero-energy edge-states. This paradigm is based on three main concepts: i) the identification of the topological phase transition within a family of structurally related polymers, ii) the chemical synthesis of the appropriate molecular precursor, and iii) the atomically precise polymerization by on-surface synthesis.

The concept of a structure related topological phase transition and the corresponding bandgap closure is exemplified in Fig 1A for the structure family of ethynylene-bridged polyacene polymers. Both DFT and tight-binding calculations shows a monotonic decrease of the HOMO-LUMO gap of the polyacene monomer as a function of the acene size (where *n* denotes the number of benzene rings of the polyacene) (*24*) (see also Fig. S1), as expected from the increasing delocalization of the frontier orbitals. However, the band gap of the ethynylene-bridged polymers shows a distinctly non-monotonic behavior, which is characterized by the closure of the bandgap for the pentacene-polymer (*n*=5) and its reopening of the gap for larger polyacenes (n>5). The determination of the Zak phase reveals that the closure of the bandgap is associated with a transition of the topological class of the polymers from trivial $Z_2$=0 to non-trivial $Z_2$=1 (see SI and Fig. S2-4 for details and definition of the topological invariant $Z_2$ in this case). The sequence of the ethynylene-bridged acene polymers can therefore be understood as representing a topological phase transition in the discrete variable n denoting the polyacene size. According to the bulk-boundary correspondence such a topological phase transition must come along with a metallic, i.e. zero bandgap interface. In the present case, our calculations show that this interface corresponds to a specific structure, which is the ethynylene-bridged pentacene polymer. Moreover, this polymer features the characteristic in-gap zero-energy edge-states as the consequence of the non-trivial topological phase.

This topological transition can also be tentatively understood in terms of the dimerized PA chain described by the SSH model (*25*). Here, the metallic phase is characterized by an equivalence of the intra-cell $t_1$ and the inter-cell $t_2$ couplings of the frontier orbitals, as the bandgap $E_g$ corresponds to twice the absolute difference of the two couplings, i.e. $E_g \sim 2|t_1 - t_2|$. Translating our system to the PA polymer case, we can take the HOMO-LUMO gap as a measure of the intra-cell coupling, which is monotonically decreasing in the pristine acene series. Similarly, the inter-cell coupling, which is provided by the ethynylene-bridge, also decreases but with smaller rate. For small polyacenes, the HOMO-LUMO gap is large and the intra-cell coupling exceeds the inter-cell one, which means that the system is in a topologically trivial case. As the size of the acene unit increases and the HOMO-LUMO gap decreases, the value of the couplings equalizes $t_1 \sim t_2$, which results in the closure of the bandgap. Further increase of the acene size makes the inter-cell coupling overcome the intra-cell one, resulting in a change of the topological class accompanied by the appearance of the zero-energy edge-states and a further increase of the bandgap.

In order to experimentally implement the outlined strategy, we have used atomically precise on-surface synthesis to produce two different polymers of the polyacene structural family. We *ex-professo* synthesized the required molecular precursors **4BrAn** (9,10-bis(dibromomethylene)-9,10-dihydroanthracene) and **4BrPn** (6,13-bis(dibromomethylene)-6,13-dihydropentacene) (see SI for synthetic details). These precursors feature a specific proaromatic molecular backbone and bear the =CBr$_2$ functionalities to steer homocoupling through a set of chemical reactions on Au(111), as recently exemplified by us for **4BrAn**(*23*). Their periodic homocoupling gives rise to the formation of ethynylene bridged anthracene (n = 3) and pentacene (n = 5) polymers, whose structural and electronic characterization is summarized in Fig. 2.

The deposition of a submonolayer coverage of **4BrAn** or **4BrPn** on Au(111) and subsequent annealing to 500 K results in the emergence of long and defect-free polymeric wires. Residual bromine atoms detached from the precursors are fully removed from the substrate at this temperature. High resolution nc-AFM imaging with CO-functionalized tip (*26*) reveals that both polymers are planar and formed by intact anthracene (Fig. 2A) and pentacene (Fig. 2F) moieties, respectively, and linked by linear bridges (Fig. 2B and 2G). Defects within the polymer chains are scarce and, in both cases, perfect wires of up to 50 nm long are readily found. The polymerization eventually completes once both $=CBr_2$ ends are lost and passivated by residual atomic hydrogen.

In the case of the anthracene polymer, scanning tunneling spectroscopy (STS) reveals a sizable band gap of ~1.5 eV (Fig. 2C-E) and inspection of the hydrogen terminated ends of the polymer show no signatures of an in-gap topological end-state (Fig. S5). Calculated band structure (Fig. 2C) and STS maps (Fig. 2E) using density functional theory (DFT) match very well with the experimental evidence. Moreover, the calculated DFT electronic structure of a finite anthracene chain also shows the absence of end-states (Fig. S6). These findings demonstrate that the anthracene polymer is in a topologically trivial electronic phase, as predicted in Fig. 1A.

In the case of the ethynylene bridged pentacene polymer, STS shows two resonances at -150 meV and at 200 meV, which we assign to the edge of the valence (VB) and conduction bands (CB), respectively (Fig. 2I). This results in the very narrow bandgap of ~0.35 eV. We should note that the band gap value obtained from STS measurements is typically reduced by additional electron screening imposed by the proximity of a metallic surface with respect to the intrinsic band gap of the gas phase polymer (*27, 28*). Nevertheless, the proximity of the topological phase transition imposes a strong band gap renormalization in its neighborhood independently of the screening effect. Consequently, the band gap of the free-standing polymer can further reduce if it is localized closer to the phase transition. This scenario is supported by quasiparticle GW calculations of the free-standing pentacene polymer, which predict nearly metallic band gap $E_g \sim 0.05\ eV$ (for more detailed discussion of the band gap see SI).

Despite the fact that DFT simulations cannot predict correctly the magnitude of the intrinsic band gap of the polymer (*29*), they describe the character of frontiers orbitals of the VB and CB edges of both polymers very well. Indeed, the agreement between experimental and simulated dI/dV maps is excellent, validating the character of the frontiers orbitals predicted by DFT calculations (see Figures 2E and J). The LDOS of the VB edge at -150 meV shows states located over the edges of the pentacene moieties (left panel of Fig. 2J), while CB has intensity maxima over the coupling bridges and on the voids adjacent to them (right panel of Fig. 2J). Interestingly, the character of the CB STS maps of the pentacene polymer is very similar to VB STS maps of anthracene polymer and vice versa. This electronic swap plays a critical role in the understanding of the topological phase and the electronic transformation of the π-conjugated polymers, as we will later show.

Regarding the topological class of the pentacene polymer, the DFT calculation predicts the presence of in-gap zero-energy edge-states at the hydrogen terminated end of finite polymer chains (Fig. 2K and Fig. S6), thus positioning it in the non-trivial region of the structural phase diagram. Experimentally, this end-state is indeed readily observed as a strong resonance by STS exactly at the Fermi level as shown in Fig. 2L-2M, corroborating its non-trivial topological nature. The small experimental bandgap is further supported by the large extent of the end-state wave function, which is inversely related to the size of the bandgap (*17*) (see also Fig. S9).

In the case of the pentacene polymer, both DFT calculations (Fig. S1) and experiment show a finite size bandgap, whereas TB suggests a basically vanishing gap, see Fig. 1B. A closer inspection of the origins of these differences in bandgap is very instructive regarding the elucidation of the fundamental relation between the topology and the electronic transformation of π-conjugated systems of the ethynylene-bridged acene polymers. In the simplistic TB model, totally rigid bonds that do not depend on the specific electronic structure of the polymer and that do not consider structure relaxation are assumed. Therefore, we cannot expect the TB model to represent the true ground state of the system, particularly in polymeric structures which can adopt different π-conjugated forms. In this context, the electronically driven bond distortion associated with the π-conjugation mechanism becomes a crucial factor, as discussed below.

Next, we show that our concept of achieving narrow bandgap polymers by tuning their structure to the topological phase boundary can be extended to other structure families. To this end, we adopted the same on-surface synthesis strategy as before to form ethynylene-bridged bisanthene polymer on Au(111) surface, this time using the **4BrBiA** (10,10'-bis(dibromomethylene)-10$H$,10'$H$-9,9'-bianthracenylidene) precursor, belong to the fused acenes family. The deposition of a submonolayer coverage of **4BrBiA** on Au(111) and subsequent annealing to 500 K gives rise to the emergence of long and defect-free polymeric wires (Fig. 3A, B). As illustrated by high resolution AFM imaging with CO-tip the polymers are planar and formed by bisanthene moieties linked by linear bridges, i.e. we keep the same bridge structure, but change to a larger repeating unit.

Again, the excellent agreement between experimental STS measurements and DFT calculations allows the unambiguous determination of the bulk electronic structure of the polymer (Fig. 3C). Experimental STS spectra recorded over the polymers on the Au(111) surface show two frontier resonances very close to the Fermi level, at - 75meV (edge of VB) and 200 meV (edge of CB) (Fig. 3D) giving a very narrow bandgap of ~0.3 eV. The dI/dV map at the VB edge shows maxima over the borders of the bisanthene units, whereas the CB edge displays a bowtie shape over the ethynylene links and adjacent voids, and some states over the bisanthene lateral borders (Fig. 3E). Interestingly, the dI/dV maps of the VB and CB edges have very similar symmetry as the pentacene polymer.

Both TB and DFT analysis of the ethynylene-bridged fused acenes family reveals that the topological phase transition with a nearly vanishing gap is expected for $n$=3, i.e. for the bisanthene polymer (Fig. S7 and S1). Its non-trivial topological quantum nature is independently confirmed by the presence of the zero-energy end-states in both experiment (Figs. 3H, I) and DFT calculations (Figs. 3G and S6).

At this stage, it is important to realize that polymers described here may adopt two extreme and distinct arrangements of the π-conjugation, either the bridge being an alkyne (ethynylene) and the acene fully aromatic or the bridge being a cumulene and the acene adopting a quinoid structure (see Fig. 1A and see also Fig. S8). In combination with the small gap this raises the question of a Jahn-Teller or Peierls type structural instability forcing an opening of the band gap. Herein, it is important to highlight that the ground state of these polymers is composed by different mixture of the two π-conjugated forms instead of their pure states (*15*) depending on the acene moiety size, with only one global minimum along the general reaction coordinate according to the DFT simulations (see also Figure S10 and underlying discussion). In principle, the existence of the single global minimum opens a possibility to tune the system to achieve the metallic state. As we mentioned above, the transition between two distinct π-conjugated resonant forms (e.g. aromatic and quinoidal), in principle, leads to a small band gap (*12-15, 30*). Thus, the question arising is whether there is a fundamental relation between the topological phase and the specific π-conjugation of the polymers?

The symmetry of the frontier orbitals seen in the dI/dV maps is not the only similarity shared by the topologically non-trivial bisanthene and pentacene polymers. High-resolution images AFM images acquired with CO-tip (*26, 31*) also show several similarities, which are not present in the case of the anthracene polymers. Namely, the AFM images of the anthracene polymer display the triple bond as a faint protrusion (see Fig. 2 and S11), revealing the presence of the ethynylene $-C \equiv C-$ configuration of the linear bridge, as recently illustrated for one-dimensional molecular wires (*23, 32*). However, in the case of the pentacene and bisanthene polymers, the triple bond signature in the bridges is inhibited as shown on Figs. 2,3 and S11, indicating increasing cumulene ($=C=C=$) character of the ethynylene bridge.

These findings highlight that the anthracene polymer adopts the ethynylene-aromatic structure of the π-conjugation, while the two topologically non-trivial polymers prefer the cumulene-like/quinoid form. Such results are also supported by a thorough analysis of the DFT simulations of the bond length variations (see explanation in SI and Fig. S12) and of the charge density isocontour plots compared with the high resolution nc-AFM images (see details in SI and Fig. S13).

The transition between aromatic/quinoid and ethynylene/cumulene character in π-conjugated polymers was rationalized by the level crossing, where HOMO/LUMO frontier orbitals of distinct symmetry swap (*30, 33*). A detailed analysis of the frontier orbitals character of all three polymers confirms this scenario (see Fig. S14). Namely, in the case of the anthracene polymer, the highest occupied wave function at K-edge of the Brillouin zone shows π-bonding character between the two C atoms forming triple bond, reinforcing the ethynylene $-C \equiv C-$ configuration. However, in the case of the pentacene and bisanthene polymer, the wave function with this symmetry is found as the lowest unoccupied orbital. This rationalizes the weakening of the triple bond. This effect, together with a relatively small energy difference between the two π-conjugated forms facilitates the promotion of the cumulene-quinoid over ethynylene-aromatic form. Importantly, the level crossing of the frontier orbitals accompanying the electronic π-conjugated transition changes the nodal character of the frontier wave functions forming the conduction band. Consequently, this gives rise to a change of the Zak phase identifying the transition of the topological phase (see also Figure S3 and underlying discussion in SI). Thus, we conclude that there is a direct relation between the topological and the π-conjugated transition.

It has been demonstrated that the aromatic character of acene monomers tends to π-electron confinement within the rings while the quinoid character promotes delocalization of π-electrons along the polymer (*17*). Therefore, the polymers consisting of the ethynylene bridge and acene monomers open playground for the competition between π-electron confinement within acene rings and delocalization along the chain (*34*). For small acene monomers with large band gap, the ground state aromatic form has much lower energy than the quinoid form. Under such circumstances, the polymers prefer to adopt the ethynylene-aromatic π-conjugation form as the ground state. However, with increasing size of the acene monomers, the energy difference between the aromatic and quinoid structure becomes small, lowering the energy penalty for promoting the cumulene/quinoid π-conjugation form along the polymer. At a certain size of the acene monomer, the enhanced contribution of the cumulene/quinoid π-conjugation in the ground state structure of the polymer causes the level crossing and the polymers transform into the topologically non-trivial phase. The situation when two frontier orbitals involved in the crossing level mechanism become degenerated corresponds exactly to the topological phase transition featuring the metallic character. In the topologically trivial phase the enhancement of the cumulene/quinoid π-conjugation leads to decrease of the band gap value, whereas in the topologically non-trivial phase this causes its increase and the appearance of the zero-energy edge-states within the gap. It is evident, that the size of the bandgap is determined by the proximity to the topological phase transition, i.e. the variant of the crossing level with degenerated frontier orbitals. The relation between the concept of the topological phase and distinct π-conjugation forms provides a bridge between the two distinct worlds of condensed matter physics and organic chemistry.

The on-surface chemical protocols presented here open new avenues for engineering novel π-conjugated polymers exhibiting unique electronic properties featuring very narrow bandgaps and in-gap zero-energy edge-states in the topological non-trivial quantum phase. We envision that these polymers could find applications in a wide variety of areas including molecular electronics, optoelectronics, organic solar cells and quantum information technology. Notably, the very low bandgap implies a direct potential use in transparent and flexible electrodes for organic optoelectronic devices (*35*). In addition, the expression of non-trivial topological quantum phase featuring the topologically protected in-gap edge-states is called to play a major role in future quantum technology organic devices like the recently proposed topological quantum memory (*36*).

Furthermore, we anticipate that the band gap of the polymer could be further modified by using the fact that the closer the structure to the crossing level/topology phase transition, the smaller the band gap becomes. Thus, exploiting the conceptual framework described here, the band gap could be tuned by a clever design of polymers with side groups (*5-8*), i.e. incorporation of heteroatoms in acene monomers or exploring ladder type topologies, targeting to synthesize purely metallic polymers. Such metallic nanomaterials would be highly desirable in polymeric science since they would provide intrinsic metallic conductivity, i.e. metallic character without chemical or electrochemical doping (*37*). In addition, intrinsic conducting polymers could revive efforts in condensed matter physics such as the quest for high Tc organic superconductors (*38*), the expression of Tomonaga-Luttinger liquid behavior in polymers, or the deeper understanding of the metal-to-insulator Mott transition in organic nanomaterials (*37*).

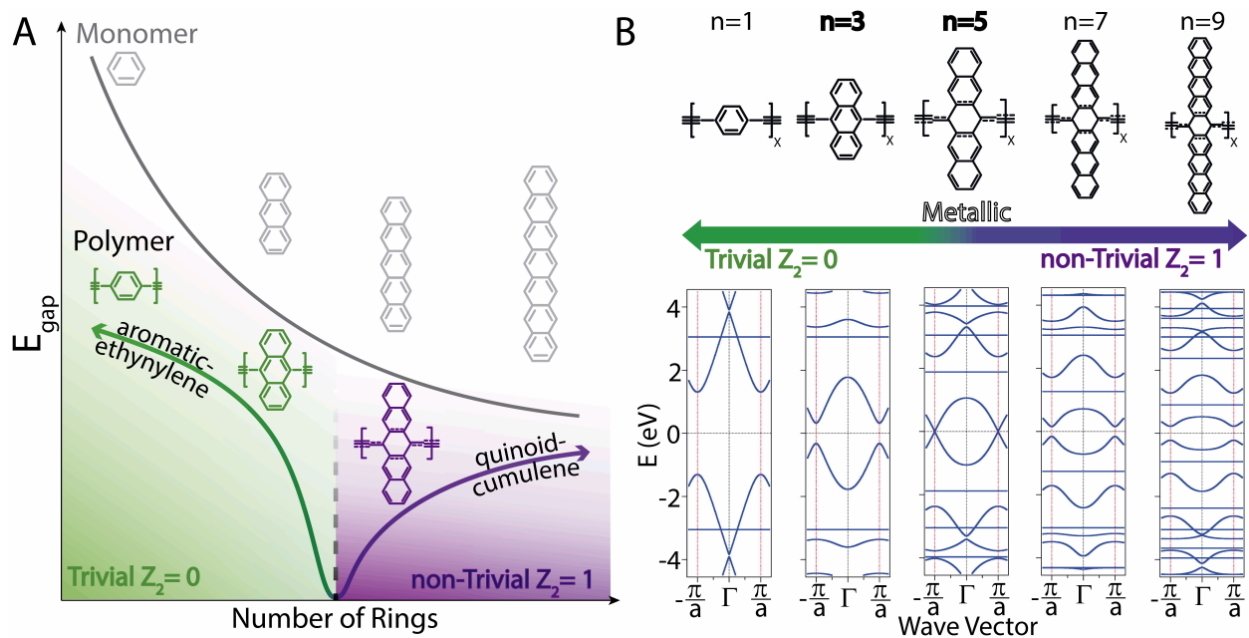

**Fig. 1**. **(A)** Schematic evolution of the bandgap of acene monomers and ethynylene-bridged acene polymers with increasing size of acene monomer. The latter case shows a phase transition between trivial and non-trivial topological classes accompanied by transformation of the pi-conjugation of the polymer. The band gap evolution with size of acenes for both monomers and polymers obtained from DFT calculations is shown in Figure S1. **(B)** Band diagram (E vs. wave vector) for the ethynylene linked acene polymers obtained from TB calculations revealing strong renormalization of the band gap near the topological phase transition.

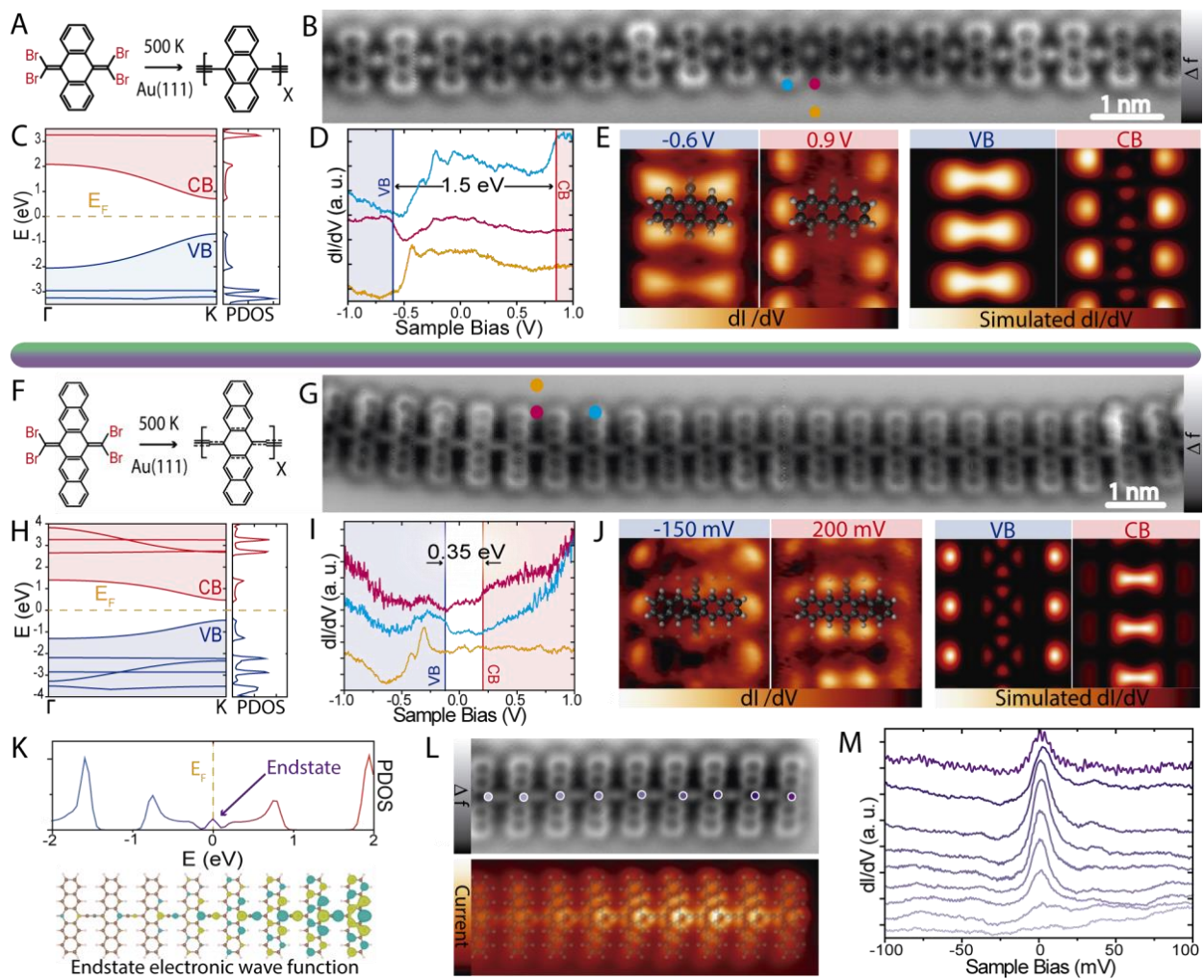

**Fig. 2.** Comparison of the experimental and theoretical results for anthracene (n=3) and pentacene (n=5) ethynylene-bridged polymers. (**A**) Chemical sketch of the resulting polymer for n=3. (**B**) nc-AFM image of the n=3 polymer. (**C**) Bulk calculated band structure and PDOS. (**D**) Experimental determination of $E_{gap}$ with STS acquired at the positions depicted in (**B**). (**E**) Constant current dI/dV maps acquired at the approximate energies of the VB and CB (left) with the corresponding simulated maps (right). (**F-J**) Same as (**A-F**) for the n=5 case (**K**) Calculated projected density of states for a finite chain of 15 units, where a close to zero-energy end-state emerges as a consequence of the non-trivial topology. (**L**) Constant height nc-AFM (top) and STM (bottom) images of an -H terminated pentacene polymer, highlighting the fading away of the end-state. (**M**) STS spectra along the termination (position depicted in **K**) showing the vanishing of the end-state. All imaging parameters are provided in the supplementary materials.

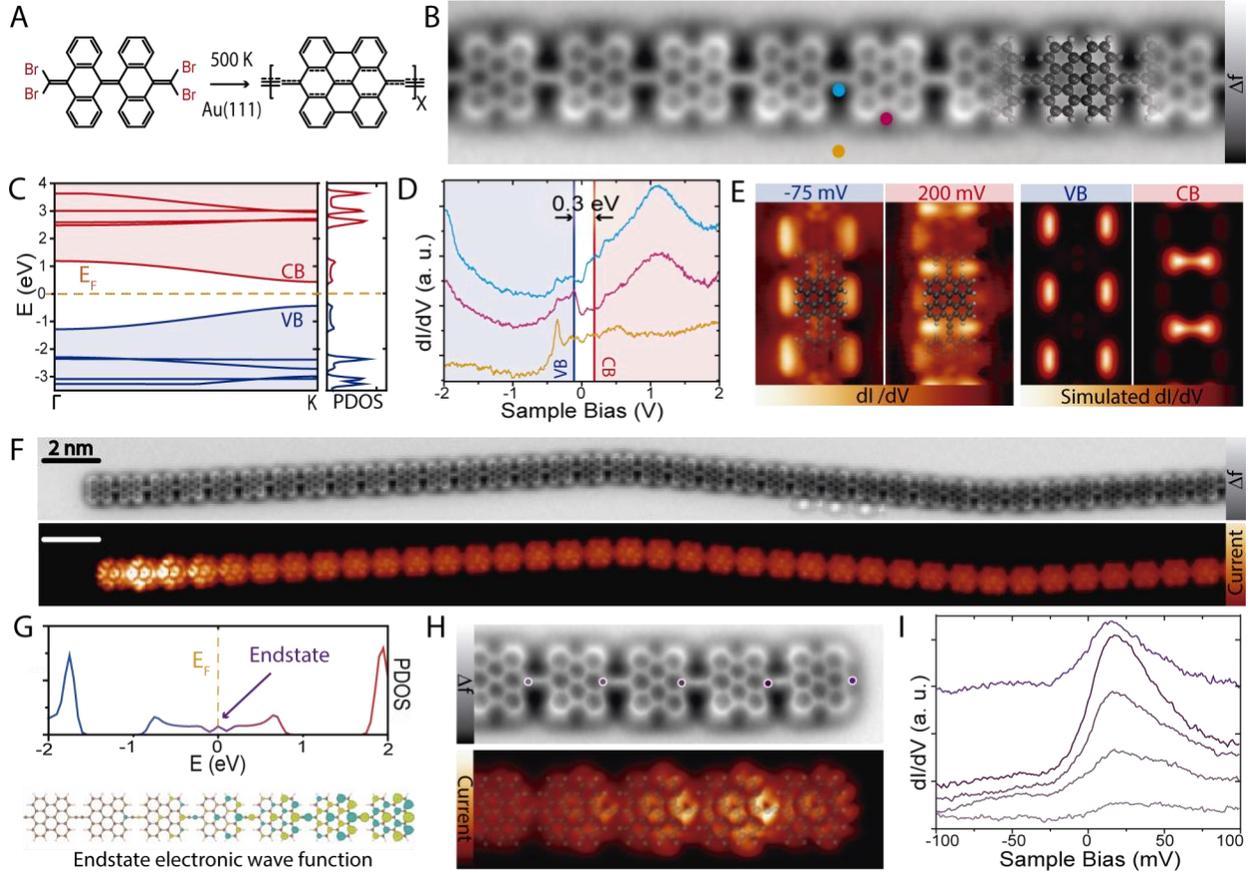

**Fig. 3**. Experimental and theoretical results for ethynylene bridged bisanthene topologically non-trivial polymer. **(A)** Reaction pathway on Au(111). **(B)** nc-AFM image of the resulting polymer with the superimposed model. **(C)** Bulk calculated band structure and PDOS. **(D)** Experimental determination of $E_{gap}$ with STS with the position highlighted in (B). **(E)** Constant current dI/dV maps acquired at the approximate energies of the VB and CB (left) with the corresponding simulated maps (right). **(F)** Large-scale constant height nc-AFM and STM images showing the end-state resulting from the non-trivial topological character of the polymer. **(G)** Calculated projected density of states for a finite chain of 15 units, where a close to zero-energy end-state emerges as a consequence of the non-trivial topology. **(H)** Constant height nc-AFM (top) and STM (bottom) images of a -H terminated polymer, highlighting the fading away of the end-state. **(I)** STS spectra along the termination showing the existence and distribution of the end-state. All imaging parameters are provided in the supplementary materials.

**Acknowledgments:**
Work supported by the ERC Consolidator Grant ELECNANO (nº 766555), the EC FP7-PEOPLE-2011-COFUND AMAROUT II, the Spanish Ramón y Cajal programme (nº RYC-2012-11133), the Spanish Ministerio de Economía y Competitividad (projects FIS 2013-40667-P, FIS 2015-67287-P), and the Comunidad de Madrid (projects MAD2D, NANOFRONTMAG (S2013/MT-2850)). Support from the European Research Council (ERC-320441-Chirallcarbon), the MINECO of Spain (projects CTQ2017-83531-R and CTQ2016-81911-REDT) and the Comunidad de Madrid (PHOTOCARBON project S2013/MIT-2841) is also acknowledged. IMDEA Nanociencia thanks support from the "Severo Ochoa" Programme for Centers of Excellence in R&D (MINECO, Grant SEV-2016-0686). P.J. acknowledges support from Praemium Academie of the Academy of Science of the Czech Republic, MEYS LM2015087 and GACR 18-09914S and Operational Programme Research, Development and Education financed by European Structural and Investment Funds and the Czech Ministry of Education, Youth and Sports (Project No. CZ.02.1.01/0.0/0.0/16_019/0000754).